\documentstyle[pra,epsf,aps,twocolumn]{revtex}

\begin{document}
\title{Relations Between Markov Processes Via Local Time
and Coordinate Transformations}
\author{A. Pelster and H. Kleinert\thanks{
Email: pelster@physik.fu-berlin.de,
kleinert@physik.fu-berlin.de;
 URL: http://www.physik.fu-berlin.de/\~{}kleinert; Phone/Fax: 0049/30/8383034.
}
}
\address{Institut f\"ur Theoretische Physik,
Freie Universit\"at Berlin, Arnimallee 14, 14195 Berlin, Germany}
\date{\today}
\maketitle
\begin{abstract}
The Duru-Kleinert (DK) method of solving unknown path integrals
of quantum mechanical systems by relating them
to known ones
does not apply to Markov processes since
the DK-transform of a Fokker-Planck equation
is in general not a Fokker-Planck equation.
In this note, we present a significant
modification of the
method, based  again on
a combination
of
path-dependent
time
and
coordinate
transformations,
to obtain
such relations after all.
As an application we express
unknown Green functions for
a one-parameter family of Markov processes
in terms of the known one for the Schenzle-Brand process.
\\
\end{abstract}
\noindent
{\bf 1.}~The stochastic theory of Markov processes \cite{Stratonovich,Kampen}
explains
many phenomena where fluctuations play a significant role.
Prominent examples are provided by
the emergence of self-organization \cite{Haken1},
the occurrence of quantum dissipation \cite{Weiss} and the appearance
of stochastic resonance \cite{Haengi}. In accordance with their
wide
range of applicability there exist various powerful solution methods
for Markov processes based either on the global characterization
of the probability evolution by the Onsager-Machlup
path integral
or on its local equivalent, the
Fokker-Planck equation.
Examples for the former are
the small noise
expansion and the
adiabatic elimination procedure of fast random variables,
for the latter the eigenfunction expansion and
the continued-fraction method applied to
periodically driven systems
\cite{Risken}.

In this note we demonstrate the use of
path-dependent time transformations,
 which have become a powerful
tool for solving quantum mechanical problems
since Duru and Kleinert's original
work on the path integral of the hydrogen atom
\cite{Duru1,Kleinert}.
In  one dimension with coordinate $q$,
the crucial time transformation has the form
\begin{eqnarray}
\label{TT}
\frac{d t}{d s} = f ( q ) \, ,
\end{eqnarray}
where  $f ( q )$ is
some positive but otherwise arbitrary function.
Such a transformation
does not change the standard formulations of
quantum mechanics,
a form invariance which has recently been emphasized in
\cite{Pelster1,Pelster2}. Due to this form invariance
different
quantum
mechanical systems
can be related to each other.

Some years ago,
the DK-method
was also applied
to the stochastic theory of
Markov processes
\cite{Blanchard,Morette}, thereby relating the Fokker-Planck equation
of Markov processes to
other stochastic differential equations.
The latter had, however, an important disadvantage:
they were no longer  Fokker-Planck equations
so that the DK-transformations did not
link
different Markov processes.
This defect will be eliminated in the sequel
by a significant modification of the
DK-method.

{}~\\
{\bf 2.}~Consider a one-dimensional Markov process of
a single random variable $x$,
whose conditional probability density
$P(x,x_0;t)$ possesses the initial condition
\begin{eqnarray}
P ( x , x_0 ; 0 ) = \delta ( x - x_0 )
\end{eqnarray}
and
obeys
the Fokker-Planck
equation for $t > 0$
\begin{equation}
\partial_t P(x,x_0;t)= \hat{H} ( x ) P ( x , x_0 ; t ) \, .
\label{}\end{equation}
Here $\hat{H} ( x )$ denotes the infinitesimal time evolution operator
\begin{eqnarray}
 \hat{H} ( x ) \bullet =
-  \frac{\partial}{\partial x}
\left[ K ( x ) \bullet \right]
+ \frac{1}{2} \frac{\partial^2}{\partial x^2}
\left[ D ( x ) \bullet \right],
\label{EQ1B}
\end{eqnarray}
containing
the
drift coefficient
$K ( x )$ and the diffusion coefficient
$D ( x )$.
Then the Laplace transform
\begin{eqnarray}
G(x,x_0;E) = \int\limits_0^{\infty} dt e^{-Et}P(x,x_0;t)
\end{eqnarray}
represents a
fixed-energy
Green function
solving the time-independent equation
\begin{eqnarray}
&& \left[ \hat{H}^{({\rm i})} ( x ) - E^{({\rm i})} \right]
G^{({\rm i})} ( x , x_0 ; E^{({\rm i})} )
+ \delta ( x - x_0 ) = 0 \, . \label{EQ1A}
\end{eqnarray}
We have added a superscript
$({\rm i})$ to emphasize that this equation describes the initial
stochastic system
from which we depart.

It is well-known \cite{Risken} that the Fokker-Planck equation
remains form invariant under arbitrary 
invertible coordinate transformations
\begin{eqnarray}
\label{RT}
x = x ( q ) \, ,
\end{eqnarray}
where the Green function transforms as
\begin{eqnarray}
\label{P1}
G^{({\rm 1})} ( q , q_0 ; E^{({\rm i})} ) = \pm \, x' ( q )
G^{({\rm i})} ( x ( q ) , x ( q_0 ) ; E^{({\rm i})} ) \, .
\end{eqnarray}
The different signs take into account whether (\ref{RT})
is monotonously in- or decreasing.
In fact, we conclude from (\ref{EQ1B}) and (\ref{EQ1A}) that 
$G^{({\rm 1})} ( q , q_0 ; E^{({\rm i})} )$ satisfies
the Fokker-Planck equation
\begin{eqnarray}
&& \left[ \hat{H}^{({\rm 1})} ( q ) - E^{({\rm i})}
\right] G^{({\rm 1})} ( q , q_0 ; E^{({\rm i})} ) + \delta ( q - q_0 )
= 0 \label{EQ2A} ,
\end{eqnarray}
where
the new infinitesimal time evolution operator
\begin{eqnarray}
 \hat{H}^{({\rm 1})} ( q ) \bullet = -
\frac{\partial}{\partial q}
\left[ K^{({\rm 1})} ( q ) \bullet \right]
+ \frac{1}{2} \frac{\partial^2}{\partial q^2}
\left[ D^{({\rm 1})} ( q ) \bullet \right]
\label{EQ2B}
\end{eqnarray}
contains the drift and diffusion coefficients
\begin{eqnarray}
K^{({\rm 1})} ( q ) & = & \frac{1}{x' ( q )} K^{({\rm i})} ( x ( q ) ) -
\frac{x'' ( q )}{2 x'{}^3 ( q )} D^{({\rm i})} ( x ( q ) ) \, , \label{DR1}\\
D^{({\rm 1})} ( q ) & = & \frac{1}{x' {}^2( q )} D^{({\rm i})} ( x ( q ) ) \, .
\label{DI1}
\end{eqnarray}
These coordinate transformations are of standard use
in finding
unknown solutions from known ones
\cite{Haken2,Schenzle}.

Let us now supplement these transformations
by
the path-dependent time transformation (\ref{TT}).
First we  proceed in
analogy with \cite[Chap.~12]{Kleinert}, and
change the
Green function
according to
\begin{eqnarray}
G^{({\rm 1})} ( q , q_0 ; E^{({\rm i})} ) & =& f ( q ) \, 
\frac{F ( q_0 ;
E^{({\rm i})} )}{F ( q ; E^{({\rm i})} )} \nonumber\\
& & \times
G^{({\rm f})} ( q , q_0 ; E^{({\rm f})} ( E^{({\rm i})} ) ) ,
\label{P2}
\end{eqnarray}
where $F(q ; E^{({\rm i})} )$ and
$E^{({\rm f})} ( E^{({\rm i})} )$
are as yet unknown trial functions.
Applying (\ref{EQ2A}) with (\ref{EQ2B}), we find
the equation
for the final
Green function
\begin{eqnarray}
&& \left[ \hat{H}^{({\rm f})} ( q ) - E^{({\rm f})} ( E^{({\rm i})} ) +
X ( q ; E^{({\rm i})} ) \right]
G^{({\rm f})} ( q , q_0 ; E^{({\rm f})}
( E^{({\rm i})} ) ) \nonumber\\
&&+ \,  \delta ( q - q_0 )
=0 ,
\label{EQ3A}
\end{eqnarray}
with the infinitesimal time evolution operator
\begin{eqnarray}
 \hat{H}^{({\rm f})} ( q ) \bullet = -
\frac{\partial}{\partial q}
\left[ K^{({\rm f})} ( q ) \bullet \right]
+ \frac{1}{2} \frac{\partial^2}{\partial q^2}
\left[ D^{({\rm f})} ( q ) \bullet \right] \, ,
\label{EQ3B}
\end{eqnarray}
containing the transformed drift and diffusion coefficients
\begin{eqnarray}
K^{({\rm f})} ( q ) & = & f ( q ) \left[ K^{({\rm 1})} ( q ) +
\frac{F' ( q ; E^{({\rm i})} )}{F ( q ; E^{({\rm i})} )}
D^{({\rm 1})} ( q ) \right] \, , \label{DR2}\\
D^{({\rm f})} ( q ) & = & f ( q ) D^{({\rm 1})} ( q ) \label{DI2}
\end{eqnarray}
and the additional term
\begin{eqnarray}
&&X ( q ; E^{({\rm i})} ) = f ( q ) \left[  \frac{1}{2}
D^{({\rm 1})} ( q ) \frac{F'' ( q ; E^{({\rm i})} )}{F ( q ; E^{({\rm i})} )}
\right. \nonumber  \\
&& \left. + K^{({\rm 1})} ( q ) 
\frac{F' ( q ; E^{({\rm i})} )}{F ( q ; E^{({\rm i})} )} 
+ {\cal E} ( q ;E^{({\rm i})}) - E^{({\rm i})}
\right] \, ,\label{EQQ}
\end{eqnarray}
where
\begin{eqnarray}
{\cal E} ( q ;E^{({\rm i})}) =  \frac{
E^{({\rm f})}(E^{({\rm i})})}{f(q)} \, .
 \label{EQQ2}
\end{eqnarray}

The equation (\ref{EQ3A}) has the above-mentioned
defect of not being a Fokker-Planck equation, due to the presence of the
additional term $X ( q ; E^{({\rm i})} )$.
This term can, however, be removed by
choosing  any
functions
$E^{({\rm f})} ( E^{({\rm i})} )$
and 
$F ( q ; E^{({\rm i})})$
which solve the differential equation
\begin{equation}
X ( q ;
E^{({\rm i})} ) \equiv  0.
\label{EQ}
\end{equation}
Note that although this
equation is of the same complexity
as the initial Fokker-Planck equation,
only a {\em particular solution\/} is required,
so that labor will definitely be saved by our method.


An alternative procedure which avoids solving the differential equation
(\ref{EQ})
is by leaving the time transformation function $f(q)$ open,
choosing some
trial function
$F ( q ; E^{({\rm i})})$, and calculating ${\cal E}(q ;
E^{({\rm i})} )$ from (\ref{EQQ}) with (\ref{EQ}).
If this happens to be factorizable
as in (\ref{EQQ2}),
the $q$-dependent prefactor may be chosen as the transformation function
$f(q)$ and the $E^{({\rm i})}$-dependent one as the energy
function $E^{({\rm f})} ( E^{({\rm i})} )$.

In either procedure, the function  $F ( q ; E^{({\rm i})})$
is subject to an important restriction. In the limit $E^{({\rm i})} 
\rightarrow 0$ it has to satisfy
\begin{eqnarray}
\label{LI0}
\lim_{E^{({\rm i})} \rightarrow 0}
F ( q ; E^{({\rm i})} ) = 1 
\end{eqnarray}
identically in $q$,
so that the energy function
$E^{({\rm f})} ( E^{({\rm i})} )$ obeys
\begin{eqnarray}
\label{LI}
\lim_{E^{({\rm i})} \rightarrow 0} E^{({\rm f})} ( E^{({\rm i})} )
= 0 \,.
\end{eqnarray}
Only under this condition do initial and final Green functions
possess proper stationary limits
\begin{eqnarray}
p^{({\rm i})}_{\rm st} ( x ) & = & \lim_{E^{({\rm i})} \rightarrow 0}
E^{({\rm i})} G^{({\rm i})} ( x , x_0 ; E^{({\rm i})} ) \, ,
\label{ST1}\\
p^{({\rm f})}_{\rm st} ( q ) & = & \lim_{E^{({\rm f})} \rightarrow 0}
E^{({\rm f})} G^{({\rm f})} ( q , q_0 ; E^{({\rm f})} ) \, .
\label{ST2}
\end{eqnarray}
>From (\ref{P1}), (\ref{P2}) and (\ref{LI0})--(\ref{ST2})
we read off a relation between them
\begin{eqnarray}
\label{STR}
p^{({\rm i})}_{\rm st} ( x ) = \pm \, 
\left[ \frac{d E^{({\rm f})} ( 
E^{({\rm i})} )}{d E^{({\rm i})}}  \right]^{-1}_{E^{({\rm i})} = 0} \,
\frac{f ( q ( x ) )}{
x' ( q ( x ) )} \,
p^{({\rm f})}_{\rm st} ( q ( x ) ) \, ,
\end{eqnarray}
which guarantees
the normalization of the probability:
\begin{equation}
\int \, p^{({\rm i})}_{\rm st} ( x ) \, d x 
=\int \, p_{\rm st}^{({\rm f})} ( q )\, d q =1
\, .
\end{equation}
An interesting feature of the present method is
that it permits us in the stationary limit to relate
the probability distributions of two {\em arbitrary\/}
Markov processes to each other.
Given initial and final drift and diffusion coefficients
$K ( x )$ and
$D ( x )$,
we
satisfy
(\ref{DI1}) and (\ref{DI2})
by choosing the time transformation function as
\begin{eqnarray}
\label{STR1}
f ( q ) =x'{}^2 ( q )  \frac{D^{({\rm f})} ( q )}{D^{({\rm i})} 
( x ( q ) )} \, .
\end{eqnarray}
Using this together with
(\ref{DR1}), (\ref{DI1}), (\ref{DR2}), (\ref{DI2}) and (\ref{LI0}),
we obtain the desired
coordinate transformation
from the differential equation
\begin{eqnarray}
\label{STR2}
x' ( q ) = C \exp \left[ \int\limits^{x ( q )} d \tilde{x}
\frac{2 K^{({\rm i})} ( \tilde{x} )}{D^{({\rm i})} ( \tilde{x} ) }
- \int\limits^q d \tilde{q} \frac{2 K^{({\rm f})} ( \tilde{q} )}{D^{({\rm f})}
( \tilde{q} )} \right] \, ,
\end{eqnarray}
where $C$ is an integration constant.
The final stationary solution
\begin{eqnarray}
p^{({\rm f})}_{\rm st} ( q ) = \frac{N^{({\rm f})} }{D^{({\rm f})} ( q )}
\exp \left[ \int\limits^q d \tilde{q}
\frac{2 K^{({\rm f})} ( \tilde{q} )}{D^{({\rm f})} ( \tilde{q} )}
\right]
\end{eqnarray}
is then related to the initial one
\begin{eqnarray}
p^{({\rm i})}_{\rm st} ( x ) = \frac{N^{({\rm i})} }{D^{({\rm i})} ( x )}
\exp \left[ \int\limits^x d \tilde{x}
\frac{2 K^{({\rm i})} ( \tilde{x} )}{D^{({\rm i})} ( \tilde{x} )}
\right] \,
\end{eqnarray}
by (\ref{STR}), (\ref{STR1}) and (\ref{STR2}),
if the normalization constants
satisfy
\begin{eqnarray}
C = \frac{N^{({\rm i})} }{N^{({\rm f})} } 
\, \left. \frac{d E^{({\rm f})} ( E^{({\rm i})} )}{d
E^{({\rm i})} }\right|_{E^{({\rm i})} = 0} \, .
\end{eqnarray}

{}~\\
{\bf 3.}~In order to demonstrate the applicability of the new transformation
method we consider
a Markov process with a  multiplicative noise for a random variable
$x \in ( 0 , \infty )$ where the drift and the diffusion coefficient
depend on an arbitrary parameter $\alpha > 0$ as follows:
\begin{eqnarray}
\label{IM}
K^{({\rm i})} ( x ) = a^{({\rm i})} x - b^{({\rm i})} x^{2 \alpha + 1} \, ,
\hspace*{0.3cm} D^{({\rm i})} ( x ) = Q^{({\rm i})} x^{2 \alpha + 2} \, .
\end{eqnarray}
Performing a transformation of the random variable (\ref{RT})
and the time (\ref{TT}) with
\begin{eqnarray}
\label{XF1}
x ( q ) = q^{\beta} \, , \hspace*{0.5cm} f ( q ) = q^{\gamma} \, ,
\end{eqnarray}
the associated differential equation
(\ref{EQ}) with (\ref{EQQ}) is solved by a  function
\begin{eqnarray}
\label{AN}
F ( q ; E^{({\rm i})} ) = q^{\delta ( E^{({\rm i})} )}  \, ,
\end{eqnarray}
if the parameters $\beta$ and $\gamma$ are related according to
\begin{eqnarray}
\label{XF2}
2 \alpha \beta + \gamma = 0 \,.
\end{eqnarray}
The function $\delta ( E^{({\rm i})} )$ and the energy
relation $E^{({\rm f})} = E^{(\rm f)} ( E^{({\rm i})} )$
are determined by
\begin{eqnarray}
\delta ( E^{({\rm i})} ) & = & \frac{\beta}{a^{({\rm i})} }
E^{({\rm i})} \, , \label{I1} \\
E^{({\rm f})} ( E^{({\rm i})} ) & = & -
\frac{ Q^{({\rm i})} }{2 a^{({\rm i})}{}^2 } E^{({\rm i})}{}^2 +
\left[ \frac{ b^{({\rm i})} }{a^{({\rm i})} } + \frac{
Q^{({\rm i})} }{2 a^{({\rm i})} } \right] E^{({\rm i})} \, .
\label{I2}
\end{eqnarray}
Note that (\ref{AN}), (\ref{I1}) and (\ref{I2}) satisfy
the correct limits (\ref{LI0}) and (\ref{LI}).

The transformed
drift
and diffusion coefficients then follow from (\ref{DR1}), (\ref{DI1}),
(\ref{DR2}) and (\ref{DI2}) as
\begin{eqnarray}
\label{BS}
K^{({\rm f})} ( q ) = a^{({\rm f})} q - b^{({\rm f})} q^{- 2 \alpha \beta
+ 1} \, , \hspace*{0.3cm} D^{({\rm f})} ( q ) = Q^{({\rm f})} q^2,
\end{eqnarray}
where
\begin{eqnarray}
a^{({\rm f})}& = & - \frac{b^{({\rm i})} }{\beta} +
\frac{Q^{({\rm i})} }{\beta} \left[
\frac{E^{({\rm i})} }{a^{({\rm i})} } - \frac{\beta - 1}{2 \beta}
\right] \, , \label{I3}\\
b^{({\rm f})} & = & - \frac{a^{({\rm i})} }{\beta}
\, , \\
Q^{({\rm f})} & = & \frac{Q^{({\rm i})} }{\beta^2} \, .\label{I4}
\end{eqnarray}

These relations supply us with
solutions of the Fokker-Planck equation
for the
one-parameter family of Markov processes (\ref{IM})
if we specialize
\begin{eqnarray}
\label{SP}
\beta = - \frac{1}{\alpha} \, .
\end{eqnarray}
Then the final Markov process (\ref{BS}) for a random variable
$q \in ( 0 , \infty )$
coincides with the
well-understood
Schenzle-Brand process
\cite{Schenzle},
which
is a standard model
in nonlinear optics and chemical reaction
dynamics. It can be derived as an approximation to
a number
of different processes by
adiabatically eliminating fast random variables in the limit of large
external fluctuations.
For instance, the Schenzle-Brand process
describes
the electrical field near a laser threshold,
the multiplicative
noise being due to inversion fluctuations.

With standard methods \cite{Risken}, the Fokker-Planck equation
of the Schenzle-Brand process with (\ref{BS}) and (\ref{SP})
can be transformed to the Schr\"odinger
equation of the Morse oscillator. As the quantum mechanical
Green function
of this system has been explicitly
calculated from path integrals \cite{Inomata,Duru},
the Green function of the Schenzle-Brand process is known:
\begin{eqnarray}
&& G^{({\rm f})} ( q , q_0 ; E^{({\rm f})} ) =
\frac{\Gamma \left(
\frac{k^{({\rm f})} }{2} + \frac{1}{4} -
\frac{ a^{({\rm f})} }{2 Q^{({\rm f})} }  \right) }{b^{({\rm f})}
\Gamma \left( 1 + k^{({\rm f})}  \right) }\,
q^{ \frac{ a^{({\rm f})} }{Q^{({\rm f})} } - \frac{5}{2} }
\nonumber \\
&& \times q_0^{- \frac{ a^{({\rm f})} }{Q^{({\rm f})}  } - \frac{1}{2} }
\exp \left[ \frac{ b^{({\rm f})} ( q_0^2 - q^2 )}{ 2 Q^{({\rm f})} } \right]
\Bigg\{ W_{\frac{1}{4} +
\frac{ a^{({\rm f})} }{2 Q^{({\rm f})}  } ,
\frac{k^{({\rm f})} }{2} }
\left( \frac{b^{({\rm f})} q^2}{Q^{({\rm f})}   }  \right) \nonumber\\
\label{G1}
&& \times M_{\frac{1}{4} +
\frac{ a^{({\rm f})} }{2 Q^{({\rm f})}  } ,
\frac{k^{({\rm f})} }{2} }
\left( \frac{b^{({\rm f})} q_0^2}{Q^{({\rm f})}  } \right)
\Theta ( q - q_0 ) +
\left( q \leftrightarrow q_0 \right) \Bigg\} \, , \label{BBSS}
\end{eqnarray}
where $k^{({\rm f})} $ denotes the abbreviation
\begin{eqnarray}
\label{K}
k^{({\rm f})}  =
\sqrt{ \left( \frac{ a^{({\rm f})} }{Q^{({\rm f})}} - \frac{1}{2}
\right)^2
+ \frac{2 E^{({\rm f})} }{Q^{({\rm f})} } }\, .
\end{eqnarray}
With the help
of the transformation formulas (\ref{P1}), (\ref{P2}),
and taking into account (\ref{XF1})--(\ref{SP}),
we find from this the unknown Green function of the
initial Markov processes (\ref{IM}):
\begin{eqnarray}
&& G^{({\rm i})} ( x , x_0 ; E^{({\rm i})} ) =
\frac{\Gamma \left(
\frac{ E^{({\rm i})} }{ 2 \alpha a^{({\rm i})} } \right) }{a^{({\rm i})}
\Gamma \left( 1 + k^{({\rm i})} \right) }
\,x^{ - \alpha k^{({\rm i})} - \alpha - 1 }
\nonumber \\
&& \times x_0^{\alpha k^{({\rm i})} + \alpha }
\exp \left[ \frac{ a^{({\rm i})} }{ 2 \alpha Q^{({\rm i})} } \left(
\frac{1}{x_0^{2\alpha}} - \frac{1}{x^{2\alpha}} \right) \right]
\nonumber \\
&& \times
\Bigg\{ \Theta ( x_0 - x )
W_{\frac{k^{({\rm i})} }{2} +
\frac{1}{2}
- \frac{E^{({\rm i})} }{2 \alpha a^{({\rm i})} } ,
\frac{ k^{({\rm i})} }{2 } }
\left( \frac{a^{({\rm i})} }{\alpha Q^{({\rm i})}  x^{2\alpha} } \right)
\nonumber \\
&& \times
M_{\frac{k^{({\rm i})} }{2}
+ \frac{1}{2}
- \frac{E^{({\rm i})} }{2 \alpha a^{({\rm i})} } ,
\frac{ k^{({\rm i})} }{2} }
\left( \frac{a^{({\rm i})} }{\alpha Q^{({\rm i})}  x_0^{2\alpha}} \right)
\label{G2}
+ \left(
x \leftrightarrow x_0 \right) \Bigg\},
\end{eqnarray}
with
\begin{eqnarray}
k^{({\rm i})} =
\frac{b^{({\rm i})} }{\alpha Q^{({\rm i})} } + \frac{1}{2 \alpha}.
\end{eqnarray}
Furthermore we obtain from (\ref{ST2}), (\ref{BBSS}), (\ref{K}) the stationary
solution of the Schenzle-Brand process (\ref{BS}), (\ref{SP})
\begin{eqnarray}
p^{({\rm f})}_{\rm st} ( q ) = \frac{ 2 \left(
\frac{b^{({\rm f})} }{Q^{({\rm f})} }
\right)^{\frac{a^{({\rm f})} }{Q^{({\rm f})} }- \frac{1}{2} } }{
\Gamma \left( \frac{a^{({\rm f})} }{Q^{({\rm f})} }- \frac{1}{2}
\right) }
q^{ 2 \frac{a^{({\rm f})} }{Q^{({\rm f})} } - 2 }
\exp \left[ - \frac{b^{({\rm f})} }{Q^{({\rm f})} } q^2 \right] \, ,
\end{eqnarray}
which is mapped via (\ref{STR}), (\ref{XF1})--(\ref{SP}) to the 
initial Markov process (\ref{IM}), yielding
\begin{eqnarray}
&& \!\!\!\!\!\!\!\!\!\!p^{({\rm i})}_{\rm st} ( x ) = \frac{2 \alpha \left(
\frac{a^{({\rm i})} }{\alpha Q^{({\rm i})} } \right)^{
\frac{b^{({\rm i})} }{ \alpha Q^{({\rm i})} }+ \frac{1}{2 \alpha} + 1 } }{
\Gamma \left(
\frac{b^{({\rm i})} }{ \alpha Q^{({\rm i})} } + \frac{1}{2 \alpha} + 1
\right) } \nonumber \\
&& \times
x^{- 2 \frac{b^{({\rm i})} }{Q^{({\rm i})} } - 2 \alpha - 2}
\exp \left[ - \frac{a^{({\rm i})} }{\alpha Q^{({\rm i})} x^{2 \alpha} }
\right] \, .
\end{eqnarray}

{}~The analytic properties of the Green functions
$G^{({\rm i})} ( x , x_0 ; E^{({\rm i})} )$ and
$G^{({\rm f})} ( q , q_0 ; E^{({\rm f})} )$ in
the energies $E^{({\rm i})}$ and $E^{({\rm f})}$ determine the
spectra
of the infinitesimal
time evolution operators $\hat{H}^{({\rm i})} ( x )$
and $\hat{H}^{({\rm f})} ( q )$ \cite[Chap.~9]{Kleinert}. From
(\ref{G1})--(\ref{G2}) we deduce that the initial multiplicative
process (\ref{IM}) has only a discrete spectrum, whereas the
final one (\ref{BS}), (\ref{SP}) contains both a discrete and a
continuous branch. Such differences between
spectral types
were encountered before in quantum mechanical DK-transformations:
the hydrogen
atom has discrete and continuous states whereas
the DK-equivalent oscillator has only discrete states 
\cite{Duru1,Kleinert}.
The spectra are usually related by a
Sommerfeld-Watson
transformation
of the Green functions \cite[Chap.~14]{Kleinert}, \cite{Mustapic}.

The discrete levels closest to zero are in a
one-to-one correspondence \cite{Pelster1,Zeile}.
For stochastic systems,
these levels
rule the approach of the conditional probability density to
its stationary limit.
In our example the associated
poles of the
initial Green function (\ref{G2})
\begin{eqnarray}
E^{({\rm i})}_n = - 2 \alpha a^{({\rm i})} n \, , \hspace*{0.5cm}
n = 0 , 1 , \ldots
\end{eqnarray}
are mapped to the corresponding final ones of (\ref{G1}), (\ref{K})
\begin{eqnarray}
E^{({\rm f})}_n = 2 Q^{({\rm f})} n^2 + ( Q^{({\rm f})} - 2 a^{({\rm f})}
) n
\end{eqnarray}
by
\begin{eqnarray}
\label{MAP}
E^{({\rm f})}_n = E^{({\rm f})} ( E^{({\rm i})}_n ) \, ,
\end{eqnarray}
as long as $n$ is bounded by
\begin{eqnarray}
n\le \frac{a^{({\rm f})} }{2 Q^{({\rm f})} } - \frac{1}{4} .
\end{eqnarray}
The evaluation of (\ref{MAP}) requires
the use of
relations (\ref{I2}) and (\ref{I3})--(\ref{SP}).

{}~\\
{\bf 4.}~In summary, we have
shown that a combination of local time and coordinate
transformations
opens new possibilities
of relating different
Markov processes.
By extending the method to several
random variables, we expect many useful applications.
Furthermore we hope that this method might help to solve
non-Markov processes \cite{Haenggi2}.

Let us finally mention that the local time transformation (\ref{TT})
is nonholonomic in spacetime, i.e., it
carries a flat spacetime into a spacetime with nonzero torsion and
curvature \cite{Pelster1,Pelster2}.
If we allow for purely spatial nonholonomic changes of coordinates,
we can reach also
spacetime geometries with spatial curvature and torsion.
This will enable us
to describe
technically relevant
diffusion processes in crystals with defects  
\cite{Kleinert2,Bausch1,Shabanov,Bausch2}.
\end{document}